\documentclass{article}
\usepackage[utf8]{inputenc}

\newcommand{\eat}[1]{}

\usepackage{amsmath, amssymb, amsfonts}

\usepackage{booktabs} 

\usepackage{graphicx}
\usepackage{amssymb}
\usepackage{amsmath}
\usepackage{graphicx}
\usepackage{latexsym}
\usepackage{epstopdf}
\usepackage{multirow}
\usepackage[vlined,ruled,commentsnumbered]{algorithm2e}
\SetEndCharOfAlgoLine{}
\usepackage{caption}
\usepackage{subcaption}
\usepackage{datetime}
\usepackage{paralist}
\setdefaultleftmargin{1em}{}{}{}{.5em}{.5em}

\usepackage{hyperref}  
\usepackage{changepage}
\usepackage{pdfpages}

\usepackage{siunitx}
\DeclareSIUnit[group-digits=true,group-separator={,}]\GB{GB}
\DeclareSIUnit[group-digits=true,group-separator={,}]\MB{MB}
\DeclareSIUnit[group-digits=true,group-separator={,}]\sec{s}
\DeclareSIUnit[group-digits=true,group-separator={,}]\msec{ms}
\DeclareSIUnit[group-digits=true,group-separator={,}]\min{min}
\usepackage{url}
\usepackage{todonotes}
\usepackage{alltt}
\usepackage{wasysym} 
\usepackage{enumitem} 
\usepackage{listings}

\usepackage{tikz}
\usepackage{tikz-qtree}

\lstdefinelanguage{scala}{
	morekeywords={abstract,case,catch,class,def,%
		do,else,extends,false,final,finally,%
		for,if,implicit,import,match,mixin,%
		new,null,object,override,package,%
		private,protected,requires,return,sealed,%
		super,this,throw,trait,true,try,%
		type,val,var,while,with,yield},
	otherkeywords={=>,<-,<\%,<:,>:,\#,@},
	sensitive=true,
	morecomment=[l]{//},
	morecomment=[n]{/*}{*/},
	morestring=[b]",
	morestring=[b]',
	morestring=[b]"""
}

\lstdefinelanguage{hypotethical-spark}{
	morekeywords={cond,var,mult,on,additive,table,foreach,PROJECT,JOIN,SUM,BY,KEY,AS,TABLE,COND,FOREACH,code},
}

\usepackage{cleveref} 

\newcommand{\tname}[1]{\texttt{#1}}
\newcommand{\cname}[1]{\texttt{#1}}

\newcommand{\ppm}{\textsl{ppm}} 

\newcommand\assignment\rho

\let\oldnl\nl
\newcommand{\nonl}{\renewcommand{\nl}{\let\nl\oldnl}}

\newcommand{\ignore}[1]{}
\newcommand{\realignore}[1]{}

\newtheorem{theorem}{Theorem}

\newtheorem{example}[theorem]{Example}

\newcommand{\systemName}{{\tt COBRA}}

\crefname{example}{example}{examples}
\Crefname{example}{Example}{Examples}

\usepackage{paralist}

\title{\systemName : Compression via Abstraction \\of Provenance for Hypothetical Reasoning}
\author{
  Daniel Deutch\\
   \small{Tel Aviv    University}
  \and
  Yuval Moskovitch\\
   \small{Tel Aviv    University}
   \and
   Noam Rinetzky\\
   \small{Tel Aviv    University}
}
\date{}

\begin{document}

\maketitle

\begin{abstract}

\end{abstract}
Data analytics often involves hypothetical reasoning: repeatedly
modifying the data and observing the induced effect on the computation
result of a data-centric application. 
Recent work has proposed to leverage ideas from data provenance tracking towards supporting efficient hypothetical reasoning: instead
of a costly re-execution of the underlying application, one may
assign values to a pre-computed \emph{provenance expression}. A prime challenge in leveraging this approach for large-scale data and complex applications lies in the size of the provenance.
To this end, we present a framework that allows to reduce provenance size. Our approach is based on reducing the provenance granularity using abstraction.    
We propose a demonstration of \systemName, a system that allows examine the effect of the provenance compression on the anticipated analysis results. We will demonstrate the usefulness
of \systemName\ in the context of business data analysis.

\section{Introduction}

Hypothetical reasoning involves examining the effect on a query/application result of modifying its input. It is a particular type of data analytics that is of great importance to analysts aiming at achieving a better understanding of the data and applications in hand, thereby optimizing either or both.  

Recent work \cite{cidr13} has proposed to leverage ideas from data provenance tracking towards supporting efficient hypothetical reasoning. The high level idea is to instrument the data with symbolic variables, either at the cell or tuple level. Then, existing provenance models such as \cite{GKT-pods07} or \cite{AggPaper} define how these variables propagate through query evaluation, to form {\em provenance polynomials}, which  may be regarded as a symbolic representation of the query result. The polynomial construction has the property that it commutes with variable valuations  \cite{AggPaper}, i.e., that the result of applying valuations directly to the computed polynomials is guaranteed to yield the same result as that of replacing the variables with the corresponding values in the input and then re-executing the query. Importantly, the former (applying valuations to the polynomial) is typically much faster than the latter (re-running the query), and so the commutativity property serves as a correctness guarantee for the reasoning process.

A prime challenge in leveraging this approach for large-scale data and complex applications lies in the \emph{provenance size}. The instrumentation process described above often results in very large provenance polynomials. While for the purpose of generating the provenance, it is reasonable to expect a rather powerful hardware, e.g., a cluster or a cloud; this assumption cannot be made for the actual interaction with the provenance expressions as applying valuation may be performed by multiple analysts, possibly using weaker hardware. Thus, requiring each such analyst to store and manipulate large polynomials may be infeasible.


In a paper that is to appear in SIGMOD~'19~\cite{provAbs},
 we presented a framework for the reduction of provenance size for hypothetical reasoning. The framework is based on the notion of {\em abstraction}; the main idea is that 
instead of assigning a distinct variable per cell/tuple, we can often group variables together, forming an abstract ``meta-variable". By doing so, we decrease the degree of freedom for hypotheticals (because now we are forced to assign a single value to all grouped variables), but we also gain in provenance size: distinct monomials may become identical, in which case they are compactly represented by a single monomial (by summing their coefficients). Whether or not it makes sense to group variables together depends on their semantics; to enable meaningful abstraction we introduced in~\cite{provAbs} the notion of \emph{abstraction trees}, which resemble ontologies over the provenance variables to guide and restrict the allowed groupings.


We propose to demonstrate our solution, which we implemented in a system called \systemName\ (for ``COmpression using aBstRAction trees''). The system  allows examining the effect of the provenance compression on the anticipated analysis results. The framework is based on the algorithm presented in \cite{provAbs}, and designed to assist the meta-analysis determine the desired bound over the compressed provenance size and the construction of the abstraction tree. This is done by presenting the changes in the analysis query results using valuation of the compressed provenance with respect to valuation of the full provenance. In more detail, \systemName\ gets as input provenance polynomials, generated by any provenance engine. The meta-analyst provides to the system a valuation for the provenance variables, an abstraction tree and a bound over the compressed provenance size. 
Once the abstraction tree and bound are set, \systemName\ computes an abstraction over the variables. The abstraction meta-variables are then presented to the user, and she may assign values, or use default values (average of the original values) set by the system. Finally, the system illustrates the effect of the compression on the analysis results by presenting the user with the query result using the full provenance compared with the result using the compressed provenance, the resulting provenance size, and the speedup in the assignment time.


\begin{figure*}[t!]
\centering
{\footnotesize
 \begin{tabular}{c}
	\begin{tabular}{|l|l|l|}
		\multicolumn{3}{c}{\tname{Cust}}\\ \hline
		\multicolumn{1}{|c|}{\cname{ID}} &
		\multicolumn{1}{c|}{\cname{Plan}} &
		\multicolumn{1}{c|}{\cname{Zip}} 
		\\ \hline
		1       &   A & 10001\\
		2       &   F1 &10001\\
		3       &   SB1&10002 \\
		4       &   Y1 &10001\\
		5       &   V &10001\\
		6       &   E &10002\\
		7       &   SB2 &10002\\
		$\dots$ & $\dots$ & $\dots$ \\
		\hline
		
	\end{tabular}
	\begin{tabular}{cc}
		\multicolumn{2}{c}{\tname{Calls}}\\
		\begin{tabular}{|l|l|r|}
			\hline \multicolumn{1}{|c|}{\cname{CID}}   &
			\multicolumn{1}{c|}{\cname{Mo}} &  \multicolumn{1}{c|}{\cname{Dur}}
			\\ \hline
			1 & 1   & 522\\
			2 & 1   & 364\\
			3 & 1   & 779\\
			4 & 1   & 253\\
			5 & 1   & 168\\
			6 & 1   & 1044\\
			7 & 1   & 697\\
			$\dots$ & $\dots$   & $\dots$\\
			\hline
		\end{tabular}
		&
		\begin{tabular}{|l|l|r|}
			\hline \multicolumn{1}{|c|}{\cname{CID}}   &
			\multicolumn{1}{c|}{\cname{Mo}} &  \multicolumn{1}{c|}{\cname{Dur}}
			\\ \hline
			1 & 3   & 480\\
			2 & 3   & 327\\
			3 & 3   & 805\\
			4 & 3   & 290\\
			5 & 3   & 121\\
			6 & 3   & 1130\\
			7 & 3   & 671\\
			$\dots$ & $\dots$   & $\dots$\\
			\hline
		\end{tabular}
	\end{tabular}
	\\
	\\

	\begin{tabular}{cc}
		\multicolumn{2}{c}{\tname{Plans}}\\
		\begin{tabular}{|l|l|c|}
			\hline
			\multicolumn{1}{|c|}{\cname{Plan}} & \multicolumn{1}{c|}{\cname{Mo}}&\multicolumn{1}{c|}{\cname{Price}} \\
			\hline
			Plan A & 1 & 0.4\\
			Family1 (F1)& 1 & 0.35\\
			Youth1 (Y1) & 1 & 0.3\\
			Veterans (V) & 1 & 0.25\\
			Small Business1 (SB1) & 1 & 0.1\\
			Small Business2 (SB2) & 1 & 0.1\\
			Enterprise (E) & 1 &  0.05\\
			$\dots$ & $\dots$ &$\dots$\\
			\hline
		\end{tabular}
		&
		\begin{tabular}{|l|l|c|}
			\hline
			\multicolumn{1}{|c|}{\cname{Plan}} & \multicolumn{1}{c|}{\cname{Mo}}&\multicolumn{1}{c|}{\cname{Price}} \\
			\hline
			A & 3 & 0.5\\
			F1& 3 & 0.35\\
			Y1 & 3 & 0.25\\
			V & 3 & 0.2\\
			SB1 & 3 & 0.1\\
			SB2 & 3 & 0.15\\
			E & 3 &  0.05\\
			$\dots$ & $\dots$ &$\dots$\\
			\hline
		\end{tabular}
	\end{tabular}
\end{tabular}	
}
\caption{Example database}\label{database}\label{Fi:database}
\end{figure*}

We will demonstrate \systemName\ in the context of business data analysis, using the synthetic telephony company database, described below, as well as data generated by the TPC Benchmark H. We will walk the audience through the process of building the abstraction trees, and let the them interactively examine the effect of the bound on the query results, provenance size and assignment time.

\paragraph*{Related Work} 
Provenance summarization was studied in multiple contexts, e.g., for probability computation \cite{DBLP:journals/pvldb/ReS08} or explanations \cite{LeeNLG17}. The main novel aspects of the present work are: (i) the problem setting which includes the use of abstraction trees that both restrict and guide the summarization, and (ii) our novel compression algorithms and analysis that leverage the presence of such trees. Indeed, the way that we use these trees to define our optimization problem is geared towards hypothetical reasoning, where one wishes to optimize the remaining degrees of freedom for hypotheticals, and is aware of the scenarios intended to be examined.

\section{Technical Background}
\label{sec:tech}

We (informally) introduce the model underlying \systemName,
through a running example. The model and the example, as well as most of the text in this section are taken verbatim or in a shortened form from \cite{provAbs}; they appear here for completeness.

\begin{example}[Running example]\label{Ex:intro-example}
	Our running example concerns a telephony company, whose database is illustrated in  
	Figure \ref{database}. It includes a \tname{Cust} table with information about the customers (ID, calling plan and zip code); a \tname{Calls} table including the duration in minutes, totaled by month for each customer; and the \tname{Plans} table including the \emph{price per minute} (\emph{\ppm}) of every plan, where the \ppm{} may vary from month to month. The company offers several calling plans: Small business plans ($SB1$, $SB2$), enterprises plan ($E$), plans for youth ($Y1$, $Y2$, $Y3$) for families ($F1, F2$) and for veterans ($V$), as well as standard plans ($A$, $B$). Each customer is subscribed to one calling plan. 
	
	Our example query computes the revenues of the company 
	by summing the per-customer-revenue, computed by multiplying the duration of calls by the \ppm{} of 
	the customer's plan, and aggregating the result per zip code:
	
	\begin{minipage}{1.1\textwidth}
		\lstset{language=sql,basicstyle=\small\normalfont\ttfamily,escapeinside={(*}{*)}}
		\begin{lstlisting}
		SELECT Zip, SUM(Calls.Dur * Plans.Price)
		FROM Calls, Cust, Plans
		WHERE Cust.Plan = Plans.Plan
		AND Cust.ID = Calls.CID
		AND Calls.Mo = Plans.Mo
		GROUP BY Cust.Zip
		\end{lstlisting}
	\end{minipage} 
	
	An analyst working for the company may be interested in the effect of
	possible changes to the call prices on the company revenues. For example, what if the \emph{price per minute} (\emph{\ppm}) of all plans are decreased by 20\% on March? Or what if	the \ppm{} in  
	the business calling plans are increased by $10\%$?
	
\end{example}

\paragraph*{Provenance Polynomials}\systemName\ gets as input provenance polynomials. Given a set of indeterminates $X$ we use the standard notion of a polynomial over $X$ as  
a sum of monomials, where each of which is a product of indeterminates and/or rational numbers referred to as coefficients. An indeterminate may appear more than once in a monomial, in which case this number of occurrences is called its exponent. We assume that we are given a {\em multiset of such polynomials}, intuitively including all polynomials that appear in the provenance-aware result of query evaluation. 

\begin{example}\label{ex:polys}
	To support the hypothetical scenarios given in Example \ref{Ex:intro-example}, we can {\em parameterize} the (multiplicative) change in price, assigning, e.g., a distinct parameter $m_i$ to capture the change in month $i$. 
	Similarly, the variables $p_1$, $f_1$, $y_1$, $v$, $b_1$, $b_2$ and $e$ are used to parameterize the plans prices based on the plan's type: $p_1$ is used to control the changes in the price of plan $A$, $f_1$ for plan $F1$, $y_1$ for $Y1$, and $v$ for the veterans plan.
	In this example we would then get as answer to the above query, instead of a single aggregate value, symbolic provenance expressions of the form 
	\begin{align*}
		P_1={} &208.8\cdot p_1\cdot m_1+ 240\cdot p_1\cdot m_3+ 127.4\cdot f_1\cdot m_1+ \\& 114.45\cdot f_1\cdot m_3+  75.9\cdot y_1\cdot m_1+ 72.5\cdot y_1\cdot m_3+ \\&42\cdot v\cdot m_1+ 24.2\cdot v\cdot m_3 \\
		P_2={} & 77.9\cdot b_1\cdot m_1+ 80.5\cdot b_1\cdot m_3+ 52.2\cdot e\cdot m_1+ \\&56.5\cdot e\cdot m_3+ 69.7\cdot b_2\cdot m_1+ 100.65\cdot b_2\cdot m_3
	\end{align*}
%
%
\end{example}

\paragraph*{Abstraction Trees} \systemName\ reduces the provenance polynomial size so that its number of monomials is below a given threshold, while supporting maximal granularity for hypothetical reasoning. To this end, we allow the user to define \emph{abstraction trees} over the
variables, intuitively defining groups of variables which will be assigned the same values. The notion of abstraction trees is critical because determining which grouping ``makes sense" is based on their semantics.
The abstraction trees may be obtained by leveraging existing ontologies on the annotated data, in turn capturing the semantics of variables. The user may also manually construct/augment the trees based on the expected use of provenance, namely, form the trees so that variables that, based on the user experience, are expected to be assigned the same value will be located in proximity to each other in the tree.


An abstraction is then represented by a {\em cut} in the tree separating the root from all leaves. The idea is that for every node in the chosen cut, all of its descendant leaves are replaced by a single metavariable. Intuitively, such choice means that for the subsequent hypothetical reasoning scenarios, all variables below each chosen node must be assigned the same value. 

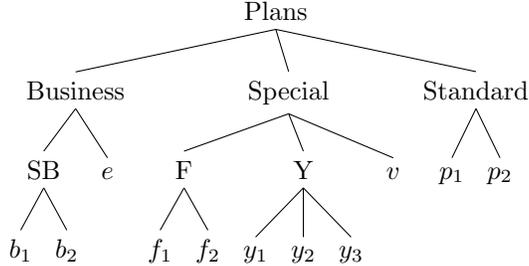
\begin{figure}
	\centering
	\begin{tikzpicture}
	\Tree [.Plans
	[.Business [.SB $b_1$ $b_2$ ]  $e$ ]
	[.Special [.F $f_1$ $f_2$ ] [.Y $y_1$ $y_2$ $y_3$ ] $v$ ]
	[.Standard $p_1$ $p_2$ ]
	]
	\end{tikzpicture}
	\caption{An 
		abstraction tree of the 
		plans variables}
	\label{fig:plansLattice}
\end{figure}

\begin{example}\label{ex:abstrees}
	In  Example~\ref{Ex:intro-example}, the plans variables may be abstracted based on their 
	type, e.g., plans for small businesses $SB1$ and $SB2$, or further abstracting all Business 
	plans, small businesses and enterprises. Abstracting the family plans using a single variable $F$, and youth plans using the variable $Y$. We may also consider using a coarse abstraction that combines all special plans (families, youth and veterans) into a single variable. Figure \ref{fig:plansLattice} depicts the resulting abstraction tree. 

%
\end{example}


\paragraph*{Optimization Problem} The problem we have studied in \cite{provAbs} is as follows: Given a provenance polynomial and abstraction tree over (subsets of) its variables, find a choice of abstraction that reduces the provenance size, while maximizing the \emph{expressiveness} of the abstraction; we next explain both measures. First, the provenance size is measured by the number of monomials in the resulting provenance polynomial. The number of monomials is indeed the dominant factor in the provenance size since each monomial is bounded by a typically small constant, independent from the database size (it may depend on the query or the number of hypothetical scenarios). As for the expressiveness of the abstraction, we aim at maximizing the degrees of freedom 
left for hypothetical analysis; naturally, every grouping limits the possible scenarios in the sense that it forces multiple variables to be assigned the same value. Consequently, we measure the expressiveness of the abstraction by the number of distinct variable names it defines. Our goal is to reduce the number of distinct {\em monomials} in the provenance, while maximizing the number of distinct {\em variables}.

\begin{example}\label{ex:validVarSet}
	Consider the abstraction tree presented in Figure \ref{fig:plansLattice}.
	The following cuts 	are possible abstractions:

$
	\begin{array}{l}
	S_1 = \{Business, Special, Standard\}\\ 
	S_2 = \{SB, e, f_1,f_2, Y, v, Standard\}\\ 
	S_3 = \{b_1, b_2, e, Special, Standard\}\\ 
	S_4 = \{SB, e, F, Y, v, p_1, p_2\}\\ 
	S_5 = \{Plans\}\\
	\end{array}
	$
\end{example}

Each choice of abstraction may entail a  ``loss" in terms of the granularity of hypothetical reasoning, in exchange for a reduction in the size of the polynomial:
	consider the polynomial $P_1$ for the revenues shown in Example \ref{ex:polys},  
	using the abstraction $S_1$ we obtain the polynomial (we use $St$ and $Sp$ as shorthand for
	Standard and Special respectively) $208.8\cdot St\cdot m_1+ 240\cdot St\cdot m_3+ 245.3\cdot Sp\cdot m_1+ 211.15\cdot Sp\cdot m_3$,
	with four different variables and four monomials,
	whereas using the abstraction $S_5$ the obtained polynomial $ 466.1\cdot Plans\cdot m_1+ 451.15\cdot Plans\cdot m_3$, consist of two monomials and three variables.

In this demonstration, we consider the case of a single abstraction tree (even in this case, a monomial may still consist of multiple variables, but the abstraction may apply to at most one of them); note that there may still be exponentially many cuts in the tree.  In this case 
the optimization problem is solvable in polynomial time complexity. In a nutshell, the algorithm traverses the abstraction tree in a bottom-up fashion, and using dynamic programming, computes an abstraction for the sub-tree rooted by each one of the inner nodes (see \cite{provAbs} for full details).

\section{System Overview}
\label{sec:system}

\begin{figure}
	\begin{center}
		\includegraphics[width=\linewidth]{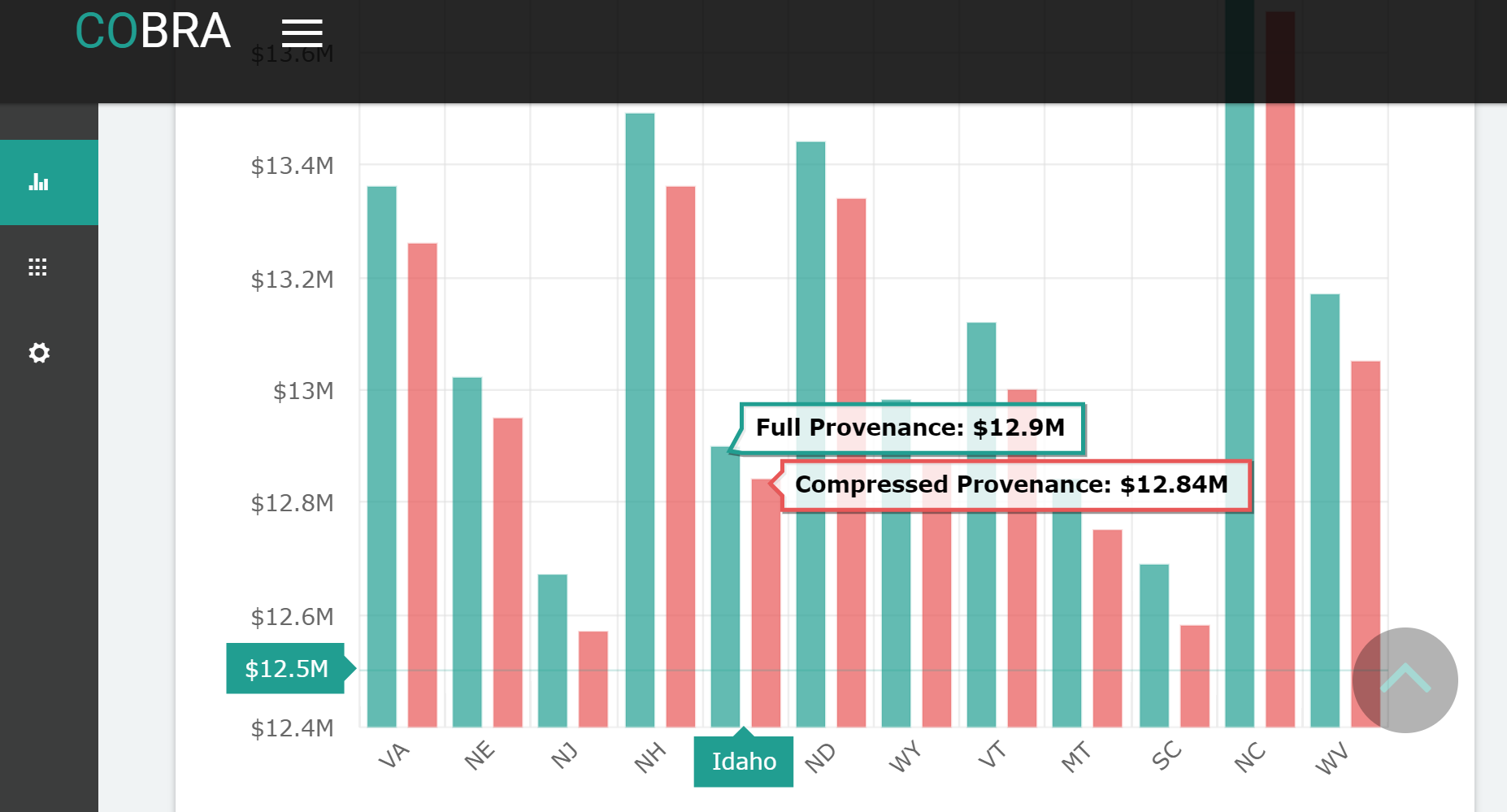}
		\caption{User Interface}\label{fig:dashboard}
	\end{center}
\end{figure}

\begin{figure}
	\begin{center}
		\includegraphics[width=\linewidth]{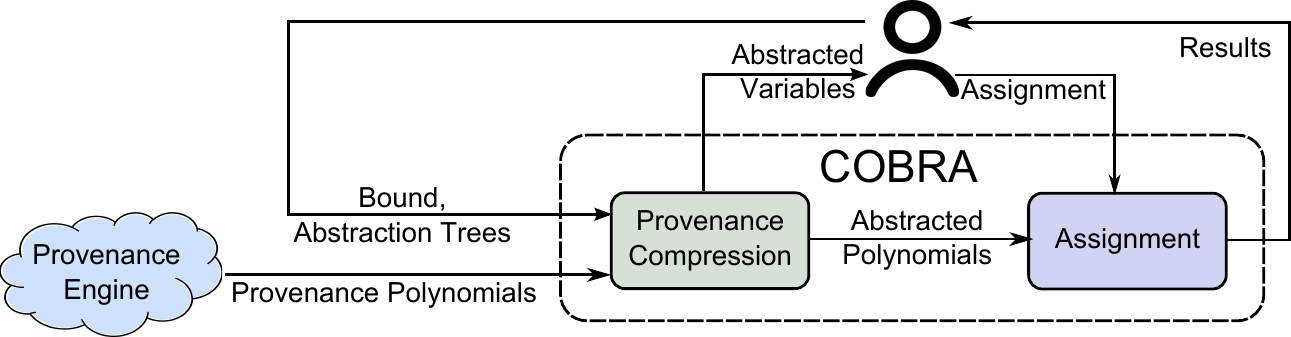}
		\caption{System Architecture}\label{fig:systemArch}
	\end{center}
\end{figure}

\systemName's back-end side is implemented in Python 3. Its front-end is written in
Angular JS framework using Bootstrap toolkit. It runs on Windows 10.  The system
architecture is depicted in Figure \ref{fig:systemArch}, and the user
interface is shown in Figure \ref{fig:dashboard}. We
next briefly explain the components of the system.

\paragraph*{Back-end} As mentioned earlier, the input to \systemName\ is set of provenance polynomials (generated by any provenance engine), default assignment to the provenance variables, a bound over the provenance size and abstraction tree (given by the user). The system then computes an optimal abstraction over the polynomials, namely, an abstraction that reduces the provenance size below the given bound while maximizing the number of variables.  This is done using the algorithm presented in \cite{provAbs}.
 Once the abstraction is generated, the user may input valuation to the compressed polynomials' variables, and the system generates the query results under the scenario given by the assignment, and presents the results to the user.

\begin{figure}
	\begin{center}
		\includegraphics[width=\linewidth]{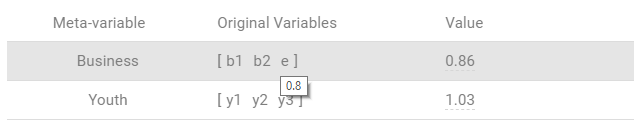}
		\caption{Meta-variables Assignment Screen}\label{fig:assignment}
	\end{center}
\end{figure}

\paragraph*{Front-end} The interaction with \systemName\ is done via a dedicated interface shown in Figure \ref{fig:dashboard}. The user is presented with the query result under a default assignment to the input provenance variables.  She can then construct the abstraction tree, and set the bound over the provenance size. 
Once the abstracted polynomials are generated, the system presents the user the abstraction variables as shown in Figure \ref{fig:assignment}. Each meta-variable in the abstraction is presented with the list of abstracted variables, each with its value in the original assignment, and a default value (average over the abstracted variables' values). The user can then modify the assigned values of the  meta-variables, and \systemName\ presents the the query result under the given assignment, showing the changes from the initial result. In addition, the system provides the user with information about the resulting provenance size and the assignment speedup using the compressed polynomials. 

\section{Demonstration Scenario}
We will demonstrate the usefulness of \systemName\  using both synthetic and real datasets.
In the first phase, we will discuss the dataset. We will use the provenance generated for the query from our running example, where the plans price was parametrized by month and plan. In addition, we will demonstrate \systemName\ in the context of TPC Benchmark H (TPC-H)~\cite{tpch}, which consists of a suite of business oriented queries. To this end, we will use the data generated by the benchmark and present a subset of its queries.

We will walk the audience through the process of building the abstraction trees, by presenting the  underlying database and the query used to generate provenance. There are multiple reasonable abstractions for each query. For instance, in our running example, if the analyst knows that the prices are usually changed uniformly during each quarter, a natural abstraction tree would consist of quarter meta-variables $q_1\ldots, q_4$, that can be used to group the monthly variables, i.e., the variables $m_1,\ldots ,m_3$ are the children of $q_1$, $m_4,\ldots,m_6$ of $q_2$ etc. The abstraction tree given in Figure \ref{fig:plansLattice} is another plausible example. We will use predefined trees for each one of the datasets.

In the second phase, we will let the audience interactively examine the effect of the bound on the query results, provenance size and assignment time. As explained in Section \ref{sec:tech}, given the provenance polynomials, abstraction tree and bound, the system computes an abstraction. Once the abstraction is computed, \systemName\ presents the user the abstraction variables with default assignment as shown in Figure \ref{fig:assignment}. We will let the user select valuations to the abstraction variables and observe the results: the changes in the analysis query results using the compressed provenance. 

Moreover, the system provides the user information about the resulting provenance size and the assignment speedup. For example, the provenance size of the polynomials generated by our running example using a database of one million customers parameterized using month variables and the leaves of the abstraction trees in Figures \ref{fig:plansLattice} is $139,260$. If we set the bound over the provenance size to $94,600$ the compressed provenance expression obtained is of size $88,620$ and the assignment speedup is $47\%$, while setting the bound to $38,600$ results in provenance polynomials of size $37,980$ and assignment speedup of $79\%$.

Finally, we will allow the audience to look ``under the hood". In particular, we will show the audience the part of the provenance polynomials, intermediate results of the algorithm and the computational sequence that lead to the resulting abstraction.

\section*{Acknowledgements}
This research has been funded by the European Research Council (ERC) under the European Union's Horizon 2020 research and innovation programme (grant agreement No. 804302), the Israeli Ministry of Science, Technology and Space, Len Blavatnik and the Blavatnik Family foundation, Blavatnik Interdisciplinary Cyber Research Center at Tel Aviv University, and the Pazy Foundation.
The contribution of Yuval Moskovitch is part of Ph.D. thesis research conducted at Tel Aviv University.

\bibliographystyle{abbrv}
\bibliography{bibliography}

\end{document}